\documentstyle[11pt]{article}
\begin{document}
\renewcommand{\theequation}{\thesection.\arabic{equation}}
\renewcommand{\refname}{References.}
\newcommand{\sect}[1]{ \section{#1} \setcounter{equation}{0} }
\newcommand{\partialslash}{\partial \! \! \! /}
\newcommand{\xslash}{x \! \! \! /}
\newcommand{\yslash}{y \! \! \! /}
\newcommand{\zslash}{z \! \! \! /}
\newcommand{\half}{\mbox{\small{$\frac{1}{2}$}}}
\newcommand{\quarter}{\mbox{\small{$\frac{1}{4}$}}}
\newcommand{\la}{\langle}
\newcommand{\ra}{\rangle}
\newcommand{\Nf}{N_{\!f}}
\title{Computation of $\beta^\prime(g_c)$ at $O(1/N^2)$ in the $O(N)$
Gross Neveu model in arbitrary dimensions.}
\author{J.A. Gracey, \\ Department of Applied Mathematics and Theoretical
Physics, \\ University of Liverpool, \\ P.O. Box 147, \\ Liverpool, \\
L69 3BX, \\ United Kingdom.}
\date{}
\maketitle
\vspace{5cm}
\noindent
{\bf Abstract.} By using the corrections to the asymptotic scaling forms of the
fields of the $O(N)$ Gross Neveu model to solve the dressed skeleton Schwinger
Dyson equations, we deduce the critical exponent corresponding to the
$\beta$-function of the model at $O(1/N^2)$.

\vspace{-16cm}
\hspace{10cm}
{\bf LTH-312}
\newpage
\sect{Introduction.}
One of the persistent problems of quantum field theories is a lack of total
knowledge of the renormalization group functions, such as the $\beta$-function,
which are important for having a precise picture of the quantum structure of
a theory. For certain models the functions can be computed to three or four
orders as a power series in a perturbative coupling constant which is
assumed to be small. However, in explicit calculations one has to work
appreciably harder to gain new information which is partly due to the
increased number of Feynman graphs one has to analyse within some
renormalization scheme such as $\overline{\mbox{MS}}$, even in theories with
the simplest of interactions. It is therefore important to develop different
techniques to give an alternative picture of the perturbative series. One
method which achieves this is the large $N$ expansion for those theories where
one has an $N$-tuplet of fundamental fields. Then $1/N$ is a small quantity for
$N$ large and this can be used as an alternative expansion parameter. Whilst a
conventional leading order analysis is relatively straightforward to carry out
for most theories it turns out that it is not useful for going to subsequent
orders in $1/N$. To obviate these difficulties methods were developed for the
$O(N)$ $\sigma$ model which were successful in solving that model at
$O(1/N^2)$, \cite{1,2}. In particular the method uses a different approach to
the conventional renormalization of the  large $N$ expansion in that one solves
the field theory precisely at its $d$-dimensional critical point, ie the
non-trivial zero of the $\beta$-function, by solving for the critical
exponents. Performing the analysis at the fixed point of the renormalization
group means that there are several simplifying features. First, the theory is
finite. Second, since $\beta(g)$ is zero the theory has a conformal symmetry
and the fields are massless. This has two consequences, one of which is that
the critical exponents of various Green's functions can be determined order by
order in $1/N$ in arbitrary dimensions, \cite{1,2}. Second, the masslessness of
the fields simplifies the Feynman integrals which occur and allows one to
compute the graphs in arbitrary dimensions, which are otherwise intractable in
the conventional (massive) large $N$ renormalization. By solving for the
exponents in this fashion one can then relate the results through an analysis
of the renormalization group equation at criticality to the critical
renormalization group functions. (See, for example, \cite{3}.) Hence, one
gains, albeit by a seemingly indirect method, information on the perturbation
series of the theory to all orders in the coupling at the order in $1/N$ one is
interested in via the $\epsilon$-expansion of the exponent. Clearly, this has
important implications for gaining a new insight into the renormalization
group functions at large orders of the coupling as well as allowing one to
check the series with explicit calculations at low orders. Furthermore, since
one calculates in arbitrary dimensions a three dimensional result will always
be determined simultaneously.

Since the earlier work of \cite{1,2} the method has been extended to models
with fermions, \cite{4,5}, supersymmetry \cite{6,7} and theories with gauge
fields [8-11]. In this paper we present the detailed evaluation of the
$\beta$-function exponent for the four-fermi theory or the $O(N)$ Gross Neveu
model, \cite{12}. The motivation for such a calculation is that first of all
knowledge of $2\lambda$ $=$ $-\, \beta^\prime(g_c)$ at $O(1/N^2)$ will mean
that the field theory will be solved completely to this order. In \cite{4} the
exponent $\eta$, which is the fermion anomalous dimension, was calculated at
$O(1/N^2)$ and more recently the vertex or $\bar{\psi}\psi$ anomalous dimension
was also computed to the same order, \cite{5}. Together with $\lambda$ one can
deduce the remaining thermodynamic exponents for this model through the
hyperscaling laws discussed in \cite{13}. Secondly, the computation of
$\lambda$-type exponents for theories with fermions as their fundamental field
is not as straightforward as the case where one deals with purely bosonic
fields. As was noted in \cite{11} there is a subtle reordering of the graphs in
the formalism and it is important to have a complete understanding of this
feature if one is to apply similar methods to deduce results in physical gauge
theories. Also to a lesser extent the methods which we had to develop to solve
the current problem, which essentially is the evaluation of massless four loop
Feynman diagrams will prove to be extremely useful in other contexts. Finally,
we are interested in going well beyond the leading order in the three
dimensional Gross Neveu model to compare estimates of various exponents from
our analytic work with Monte Carlo simulations currently being carried out,
\cite{14}. The leading order results are not precise enough to be able to
compare with the relatively low values of $N$ which are being simulated and the
$O(1/N^2)$ reults therefore must be computed.

The paper is organised as follows. In section 2, we introduce our notation and
review the leading order formalism used to compute the exponent $\lambda$ at
$O(1/N)$ which will serve as the foundation for the $O(1/N^2)$ corrections.
This formal extension is discussed in section 3 where we derive finite
consistency equations and explain the need to compute some three and four
loop Feynman graphs. The explicit evaluation of these is discussed in
sections 4 and 5 whilst the $O(1/N^2)$ corrections to a $2$-loop integral
which appears at $O(1/N)$ are derived in section 6. We conclude our
calculation in section 7 by giving an arbitrary dimensional expression for
$\lambda$ at $O(1/N^2)$ and discuss the numerical predictions deduced from it
for the three dimensional model.

\sect{Preliminaries.}
The theory we consider involves self interacting fermions $\psi^i$, $1$ $\leq$
$i$ $\leq$ $N$, where $1/N$ will be the expansion parameter for $N$ large. One
can formulate the lagrangian either by using the explicit four point
interaction or by introducing a bosonic auxiliary field, $\sigma$, which is the
version we use. The quantum theory of both are equivalent. Thus we take,
\cite{12},
\begin{equation}
L ~=~ \frac{1}{2} \bar{\psi}^i \partialslash \psi^i + \frac{1}{2} \sigma
\bar{\psi}^i\psi^i - \frac{\sigma^2}{2g^2}
\end{equation}
where $g$ is the perturbative coupling constant which is dimensionless in two
dimensions. The aim will be to calculate the $O(1/N^2)$ corrections to the
$\beta$-function and we note that the three loop structure of this has
already been calculated perturbatively in dimensional regularization using the
$\overline{\mbox{MS}}$ scheme, \cite{A,15,16,17}, as
\begin{equation}
\beta(g) ~=~ (d-2)g - (N-2)g^2 + (N-2)g^3 + \quarter (N-2)(N-7) g^4
\end{equation}
where the coupling constant in (2.2) is related to that of (2.1) by a factor
$2\pi$ which we omit here since it will play a totally passive role in the
rest of the discussion. It is important to note that in carrying out
perturbative calculations with dimensional regularization that (2.2) is what
one determines as the $\beta$-function in $d$-dimensions prior to setting
$d$ $=$ $2$ to obtain the renormalization group functions in the original
dimension. There, of course, the theory is asymptotically free. However, the
$d$-dimensional $\beta$-function (2.2) can be viewed from a different point
of view for the large $N$ critical point analysis of the present work. For
instance, when $d$ $>$ $2$ which is the case we will deal with for the rest of
the paper, there exists a non-trivial zero of the $\beta$-function at a
value $g_c$ given by
\begin{equation}
g_c ~ \sim ~ \frac{\epsilon}{(N-2)}
\end{equation}
at leading order in large $N$ where the corrections are $O(\epsilon^2)$ and
$O(1/(N-2)^2)$ and $d$ $=$ $2$ $+$ $\epsilon$. This corresponds to a phase
transition which is apparent in the explicit three dimensional work of
\cite{18,19}. Indeed similar approaches were examined in the work of \cite{20}
for the $O(N)$ $\sigma$ model. When one is in the neighbourhood of a phase
transition, it is well known that physical quantities possess certain power law
behaviour. For physical systems the power or critical exponent fundamental to
the power law totally characterizes the properties of the system. In continuum
field theory, in the neighbourhood of a phase transition, the Green's functions
also exhibit a power law structure where the critical exponent, by the
universality principle, has certain properties, \cite{3}.  For instance, it
depends only on the spacetime dimension and any internal parameters of the
underlying field theory. More importantly, though, for our purposes, one can
solve the renormalization group equation at criticality and relate several
exponents to the fundamental functions of the renormalization group equation,
\cite{3}, which like (2.2) are ordinarily calculated order by order in
perturbation theory. In the alternative critical point approach one can compute
the exponents at several orders in $1/N$ which then gives independent
information on that critical renormalization group function. Since the location
$g_c$ is known at leading order as a function of $\epsilon$ and $1/N$, one can
undo the relations between exponent and critical renormalization group function
to deduce the coefficients appearing in the perturbative series. Clearly this
is a powerful alternative method of computing, say, $\beta$-functions.

For this paper, we extend the earlier work of \cite{11} which was based on the
pioneering techniques developed for the bosonic $\sigma$ model on $S^N$,
\cite{1,2}. The physical ideas behind the method are relatively simple. In the
neighbourhood of $g_c$ the model is conformally symmetric and therefore the
Green's functions scale. To analyse the critical theory one postulates the
most general structure the Green's functions can take which is consistent with
Lorentz and conformal symmetry, \cite{1}. The critical exponents of these
scaling forms involves two pieces. One is related in the case of a
propagator to the canonical dimension of the field as defined by the fact that
the classical action with lagrangian (2.1) is a dimensionless object. Since
quantum fluctuations will always alter the canonical dimension a non-zero
anomalous dimension is appended to the canonical dimensional and it carries
the information relevant for the renormalization group functions.

To be more concrete and to fix notation, for (2.1) the scaling forms of the
propagators in coordinate space as $x$ $\rightarrow$ $0$ are, \cite{4}
\begin{equation}
\psi (x) ~ \sim ~ \frac{A\xslash}{(x^2)^\alpha} ~~~,~~~
\sigma(x) ~ \sim ~ \frac{B}{(x^2)^\beta}
\end{equation}
where
\begin{equation}
\alpha ~=~ \mu + \half \eta ~~~,~~~
\beta ~=~ 1 - \eta - \chi
\end{equation}
and $\eta$ is the fermion anomalous dimension, $\chi$ is the vertex anomalous
dimension and $d$ $=$ $2\mu$ is the spacetime dimension. Both $\eta$ and $\chi$
have been calculated at $O(1/N^2)$ within the self consistency approach,
\cite{4,5}, and are
\begin{equation}
\eta_1 ~=~ \frac{2(\mu-1)^2\Gamma(2\mu-1)}{\Gamma(2-\mu)\Gamma(\mu+1)
\Gamma^2(\mu)}
\end{equation}
\begin{equation}
\eta_2 ~=~ \frac{\eta^2_1}{2(\mu-1)^2} \left[ \frac{(\mu-1)^2}{\mu}
+ 3\mu + 4(\mu-1) + 2(\mu-1)(2\mu-1)\Psi(\mu)\right]
\end{equation}
\begin{equation}
\chi_1 ~=~ \frac{\mu\eta_1}{(\mu-1)}
\end{equation}
\begin{eqnarray}
\chi_2 &=& \frac{\mu\eta^2_1}{(\mu-1)^2} \left[ 3\mu(\mu-1)\Theta(\mu)
+ (2\mu-1)\Psi(\mu) \right. \nonumber \\
&&- ~\left. \frac{(2\mu-1)(\mu^2-\mu-1)}{(\mu-1)} \right]
\end{eqnarray}
where $\Psi(\mu)$ $=$ $\psi(2\mu-1)$ $-$ $\psi(1)$ $+$ $\psi(2-\mu)$ $-$
$\psi(\mu)$, $\Theta(\mu)$ $=$ $\psi^\prime(\mu)$ $-$ $\psi^\prime(1)$ and
$\psi(\mu)$ is the logarithmic derivative of the $\Gamma$-function. The
expression (2.6) was first derived in \cite{21} and later in [23-25] where
$\chi_1$ was also given in \cite{22}. Each expression (2.6)-(2.9) agrees with
the respective three loop perturbative results of the corresponding
renormalization group functions in $d$ $=$ $2$ $+$ $\epsilon$ dimensions
which were given in \cite{15}. The quantities $A$ and $B$ in (2.4) are the
amplitudes of $\psi$ and $\sigma$ respectively and are independent of $x$.

The method to deduce (2.6) and (2.7) is to take the ans\"{a}tze (2.4) and (2.5)
and substitute them into the skeleton Dyson equations with dressed propagators
of the $2$-point function, \cite{1,4}, which are valid for all values of the
coupling including the critical coupling. One subsequently obtains a set of
self consistent equations which represent the critical Dyson equations. Their
solution fixes $\eta_i$ which is the only unknown at the $i$th order where
$\eta$ $=$ $\sum_{i=1}^\infty \eta_i/N^i$. Further the vertex anomalous
dimension $\chi$ is determined by considering the scaling behaviour of the
$\sigma\bar{\psi}\psi$ vertex also in the critical region using a method
developed in \cite{5} which extended the earlier work of \cite{25} to
$O(1/N^2)$.

To determine the corrections to the $\beta$-function one follows the analogous
procedure used in \cite{2} and developed for models with fermion fields in
\cite{4,11}. If we set $2\lambda$ $=$ $- \, \beta^\prime(g_c)$ then the
critical slope of the $\beta$-function can be computed by considering the
corrections to the asymptotic scaling, \cite{4}, ie
\begin{eqnarray}
\psi(x) &\sim& \frac{A}{(x^2)^\alpha} \left[ 1 + A^\prime(x^2)^\lambda
\right] \nonumber \\
\sigma(x) &\sim& \frac{B}{(x^2)^\beta} \left[ 1 + B^\prime(x^2)^\lambda
\right]
\end{eqnarray}
where $A^\prime$ and $B^\prime$ are new amplitudes and $\lambda$ $=$ $\mu$
$-$ $1$ $+$ $\sum_{i=1}^\infty \lambda_i/N^i$ from (2.2). The idea then is to
compute $\lambda$ at $O(1/N^2)$ in arbitrary dimensions. Once obtained we can
use the relation between $\lambda$ and $\beta^\prime(g_c)$ to deduce $\beta(g)$
as a power series in $g$ at the same approximation in large $N$, since
knowledge of $\lambda_1$ allows us to determine the value of $g_c$ to undo the
relation.

We close the section by reviewing the method of \cite{4,11} to deduce
$\lambda_1$. As indicated we use the skeleton Dyson equations which are
illustrated in fig. 1. To deduce $\eta_1$ the equations were truncated by
including only the one loop graphs of fig. 1. However, it turns out, as we
will recall below, that for $\lambda_1$ one has to consider the additional
two loop graph of the $\sigma$ equation. The quantities $\psi^{-1}$ and
$\sigma^{-1}$ are the respective two point functions and their asymptotic
scaling forms have been deduced from (2.4) by inverting in momentum space
using the Fourier transform
\begin{equation}
\frac{1}{(x^2)^\alpha} ~=~ \frac{a(\alpha)}{2^{2\alpha}\pi^\mu} \int_k
\frac{e^{ikx}}{(k^2)^{\mu-\alpha}}
\end{equation}
where $a(\alpha)$ $=$ $\Gamma(\mu-\alpha)/\Gamma(\alpha)$. Thus as $x$
$\rightarrow$ $0$, \cite{4},
\begin{eqnarray}
\psi^{-1}(x) & \sim & \frac{r(\alpha-1)\xslash}{A(x^2)^{2\mu-\alpha+1}}
\left[ 1 - A^\prime s(\alpha-1)(x^2)^\lambda \right] \\
\sigma^{-1}(x) & \sim &\frac{p(\beta)}{B(x^2)^{2\mu-\beta}}
\left[ 1 - B^\prime q(\beta) (x^2)^\lambda \right]
\end{eqnarray}
where
\begin{eqnarray}
p(\beta) &=& \frac{a(\beta-\mu)}{\pi^{2\mu}a(\beta)} ~~~,~~~
r(\alpha) ~=~ \frac{\alpha p(\alpha)}{(\mu-\alpha)} \\
q(\beta) &=& \frac{a(\beta-\mu+\lambda)a(\beta-\lambda)}{a(\beta-\mu)a(\beta)}
{}~~~,~~~ s(\alpha) ~=~ \frac{\alpha(\alpha-\mu)q(\alpha)}{(\alpha-\mu+\lambda)
(\alpha-\lambda)} \nonumber
\end{eqnarray}
To represent the graphs of fig. 1 one merely substitutes (2.10), (2.12) and
(2.13) for the lines of each of the graphs to obtain
\begin{eqnarray}
0 &=& r(\alpha-1)[1-A^\prime s(\alpha-1)(x^2)^\lambda] + z[1+(A^\prime
+ B^\prime)(x^2)^\lambda] \\
0 &=& \frac{p(\beta)}{(x^2)^{2\mu-\beta}} [1-B^\prime q(\beta)(x^2)^\lambda]
+ \frac{Nz}{(x^2)^{2\alpha-1}}[1+2A^\prime (x^2)^\lambda] \nonumber \\
&-& \frac{Nz^2}{2(x^2)^{4\alpha+\beta-2\mu-2}} [ \Pi_1 + (\Pi_{1A}A^\prime
+ \Pi_{1B}B^\prime)(x^2)^\lambda]
\end{eqnarray}
where we have not cancelled the powers of $x^2$ in (2.10) and the quantities
$\Pi_1$, $\Pi_{1A}$ and $\Pi_{1B}$ are the values of the two loop integral
in the respective cases when there are no $(x^2)^\lambda$ contributions,
when $(x^2)^\lambda$ is included on a $\psi$ line and when it is included on
the $\sigma$ field. We have also set $z$ $=$ $A^2B$. As in \cite{2,4} the terms
of (2.15) and (2.16) involving powers of $(x^2)^\lambda$ decouple from those
which do not to leave two sets of consistency equations. One set yields
$\eta_1$ whilst the second determine $\lambda_1$. To achieve this one forms a
$2$ $\times$ $2$ matrix which has $A^\prime$ and $B^\prime$ as the basis
vectors and sets its determinant to zero to have a consistent solution. It is
the subtlety of taking this determinant which necessitates the inclusion of
$\Pi_{1B}$, \cite{4}, in (2.16) as discussed in \cite{11}. Basically when one
substitutes the leading order values for $\alpha$ and $\beta$ into the basic
functions (2.14) one finds
\begin{equation}
s(\alpha-1) ~=~ O(N) ~~,~~ r(\alpha-1) ~=~ O\left( \frac{1}{N} \right) ~~,~~
q(\beta) ~=~ O\left( \frac{1}{N} \right)
\end{equation}
Thus analysing the leading order $N$ dependence of each of the elements of the
matrix one finds that the contribution from the terms in $\sigma^{-1}$
involving $B^\prime$ are of the same order as the (finite) two loop graph
$\Pi_{1B}$. Thus it cannot be neglected and we note that explicit evaluation
gave
\begin{equation}
\Pi_{1B} ~=~ \frac{2\pi^{2\mu}}{(\mu-1)^2\Gamma^2(\mu)}
\end{equation}
Thus substituting into the equation
\begin{equation}
\det \left(
\begin{array}{cc}
- \, r(\alpha-1)s(\alpha-1) & z \\
2z & - \, \frac{p(\beta)q(\beta)}{N} - \frac{z^2}{2}\Pi_{1B} \\
\end{array}
\right) ~=~ 0
\end{equation}
one deduces
\begin{equation}
\lambda_1 ~=~ - \, (2\mu-1)\eta_1
\end{equation}
as was recorded in \cite{4}. This completes our review of the previous work in
this area and lays the foundation for the subsequent higher order calculations.

\sect{Master equation.}

In this section, we derive the formal master equation whose solution will yield
$\lambda_2$. As already indicated in the previous section this involves
truncating the Dyson equations at the next order and including the appropriate
corrections. For the moment we concentrate on the equation for $\psi$ as it has
a simpler structure compared to (2.16). The additional $O(1/N^2)$ correction we
consider is illustrated in fig. 2 and we denote it by $\Sigma$. Including it in
(2.15) we have
\begin{eqnarray}
0 &=& \frac{r(\alpha-1)}{(x^2)^{2\mu - \alpha + 1}}[ 1-A^\prime s(\alpha-1)
(x^2)^\lambda] + \frac{z m^2}{(x^2)^{\alpha+\beta-\Delta}}
[1+(A^\prime+B^\prime)(x^2)^\lambda] \nonumber \\
&+& \frac{z^2}{(x^2)^{3\alpha+2\beta-2\mu-1-2\Delta}} [ \Sigma + (A^\prime
\Sigma_A + B^\prime\Sigma_B)(x^2)^\lambda]
\end{eqnarray}
where the subscript on the corrections $\Sigma_A$ and $\Sigma_B$ correspond to
the insertion of $(x^2)^\lambda$ on either the $\psi$ or $\sigma$ lines of the
graph of fig. 2 and $\Sigma$, $\Sigma_A$ and $\Sigma_B$ are the values of the
respective integrals. There are two graphs making up $\Sigma_B$ due to the
presence of two $\sigma$ lines and each give the same contribution. For
$\Sigma_A$, there are three graphs, one of which gives a different value from
the other two where the insertion is on a line adjacent to the external vertex.
In (3.1) we have included the additional quantities $\Delta$ and $m$. The
graph $\Sigma$ arises in \cite{4} in the determination of $\eta_2$ and it is
in fact infinite which can be seen by the explicit computation using the
uniqueness method developed first in \cite{26} and later in \cite{2,27}.
Consequently, one has to introduce a regularization by shifting the exponent of
the $\sigma$ field by an infinitesimal quantity $\Delta$, ie $\beta$
$\rightarrow$ $\beta$ $-$ $\Delta$. To remove the infinities from $\Sigma$,
$\Sigma_A$ and $\Sigma_B$ one uses the counterterm available from the leading
order one loop graph. Thus formally setting
\begin{equation}
\Sigma ~=~ \frac{K}{\Delta} + \Sigma^\prime ~~~,~~~
\Sigma_{A,B} ~=~ \frac{K_{A,B}}{\Delta} + \Sigma^\prime_{A,B}
\end{equation}
in (3.1) and expanding
\begin{equation}
m ~=~ 1 ~+~ \frac{m_1}{\Delta N} ~+~ O \left( \frac{1}{N^2} \right)
\end{equation}
the divergent terms of (3.1) are set to zero minimally to obtain a finite
consistency equation ie
\begin{equation}
m_1 ~=~ - \, \frac{z_1K_A}{2} ~=~ - \, \frac{z_1 K_B}{2}
\end{equation}
which implies $K_A$ $=$ $K_B$ and this will provide a check on the explicit
calculation described later. In order to proceed to the critical region one
must exclude the $\ln x^2$ style terms which remain which is achieved by
exploiting the freedom in the definition of the vertex anomalous dimension by
setting
\begin{equation}
\chi_1 ~=~ - \, z_1 K_A ~=~ - \, z_1 K_B
\end{equation}
Agreement with (2.8) will be another check. This will leave a finite set of
equations which are valid as $x$ $\rightarrow$ $0$ which again decouples into
one which is relevant for $\eta_2$ and the other for $\lambda_2$ ie \cite{4}
\begin{equation}
0 ~=~ r(\alpha-1) + z + z^2 \Sigma^\prime
\end{equation}
from which $z_2$ can be derived and
\begin{equation}
0 ~=~ A^\prime [z-r(\alpha-1)s(\alpha-1) + z^2\Sigma^\prime_A]
+ B^\prime[z+z^2\Sigma^\prime_B]
\end{equation}
However, by analysing the $N$-dependence of each term of the $A^\prime$
coefficient of (3.7) the correction $\Sigma^\prime_A$ is $O(1/N^2)$ with
respect to $r(\alpha-1)s(\alpha-1)$ and therefore does not need to be computed
explicitly since it will contribute to $\lambda_3$ and not $\lambda_2$.

We now turn to the $\sigma$ equation. In the same way we had to consider the
higher order two loop graph of fig. 1 to deduce $\lambda_1$, we now have to
include the analogous set of graphs for the next order to determine
$\lambda_2$. As we are using graphs with dressed propagators it turns out
there are only five graphs which arise. These are illustrated in figs 3 and 4
and we have given each a label. The subscript $B$ indicates that we need only
consider the graphs where there is an $(x^2)^\lambda$ insertion on the $\sigma$
line. Again the insertions on the $\psi$ lines will be relevant for
$\lambda_3$. We have grouped the graphs which are divergent and therefore
require regularization by $\Delta$. The origin of the infinity is the same as
the vertex infinity which occurs in $\Sigma$. Indeed in fig. 3 each graph
corresponds to the usual vertex correction of the two loop graph $\Pi_1$. For
the $\lambda$ consistency equation of the $\sigma$ Dyson equation there is an
insertion of $(x^2)^\lambda$ in one of the $\sigma$ lines of the graphs. For
the case $\Pi_{3B}$, for example, the insertion in one line removes the
infinity from that vertex since the presence of the exponent $\lambda$ on the
$\sigma$ lines moves the overall exponent of that line from the value which
gives infinity. Thus the graph has a simple pole in $\Delta$ which is removed
by the same vertex counterterm as (3.4). For $\Pi_{2B}$ one of the insertions
on a $\sigma$ line makes the graph finite and we call it $\Pi_{2B2}$ and no
regularization is required. The other case, $\Pi_{2B1}$, is divergent but is
again rendered finite by (3.4) in the consistency equation. By contrast the
graphs of fig. 4 do not involve any divergent vertex subgraphs and when any
one of the $\sigma$ lines has an $(x^2)^\lambda$ insertion each graph is
completely finite. For notational convenience we define $\Pi_{5B1}$ to be the
graph with an insertion on the top $\sigma$ line and $\Pi_{5B2}$ to have an
insertion on the central $\sigma$ line. Similarly, we denote the graph of
$\Pi_{6B}$ with the bottom $\sigma$ line corrected by $\Pi_{6B1}$ and the
other case by $\Pi_{6B2}$. Therefore, there are five distinct finite
massless Feynman graphs to evaluate.

We have dealt at length with the higher order graphs which need to be included
in the corrections to (2.16). However, there are also $O(1/N^2)$ contributions
coming from $\Pi_{1B}$. For, $\lambda_1$, one considers this graph with
$\alpha$ $=$ $\mu$, $\beta$ $=$ $1$ and $\lambda$ $=$ $\mu$ $-$ $1$. These are
only the leading order values of the exponents and since we are dealing with
fields with non-zero anomalous dimensions which are $O(1/N)$ these give
contributions to $\lambda_2$ when $\Pi_{1B}$ is expanded in powers of $1/N$.
Therefore, these ought not to be neglected in deriving the master equation
for $\lambda_2$.

Rather than reproduce the analogous renormalization of the $\sigma$
consistency equation, which proceeds along the same straightforward lines as
discussed earlier, we obtain the following finite equation which includes the
corrections to (2.16)
\begin{equation}
0 ~=~ 2zA^\prime - B^\prime \left[ \frac{p(\beta)q(\beta)}{N} +
\frac{z^2}{2}\Pi_{1B} + \frac{z^3}{2}\Pi_{B2} \right]
\end{equation}
where
\begin{eqnarray}
\Pi_{B2} &=& 2\Pi_{2B1} + 2\Pi_{2B2} + 2\Pi_{3B} + 2\Pi_{4B} \nonumber \\
&-& z_1 [ 2\Pi_{5B1} + \Pi_{5B2} + 2\Pi_{6B1} + 4\Pi_{6B2} ]
\end{eqnarray}
and the prime is understood on the integrals which are divergent. We have
preempted the explicit calculation of later sections by using the fact that
$\Pi^\prime_{1A}$ $=$ $0$ in writing down (3.8) and ignoring the corrections
where there is an insertion on the $\psi$ lines of the graphs of figs. 3 and
4 since they are relevant for $\lambda_3$.

This completes the derivation of the formal consistency equations which yield
$\lambda_2$. One again sets the determinant of the $2$ $\times$ $2$ matrix
formed by $A^\prime$ and $B^\prime$ as the basis vectors in (3.7) and (3.8) to
zero, and all that remains is the evaluation of $\Pi_{1B}$ and $\Pi_{B2}$.

\sect{Computation of divergent graphs.}

In this section we discuss the computation of the divergent graphs $\Pi_{1A}$,
$\Sigma_{1A}$, $\Pi_{2B1}$ and $\Pi_{3B}$. First, though we recall the basic
tool we use for computing massless Feynman graphs, which is the uniqueness
construction first used in \cite{26} and developed for large $N$ work in
\cite{2} and other applications in \cite{27}. The basic rule for a bosonic
vertex which we require is illustrated in fig. 5 whilst that for a
$\sigma \bar{\psi} \psi$ type vertex is given in fig. 6, \cite{4}. In each
case the arbitrary exponents $\alpha_i$ and $\beta_i$ are constrained to be
their uniqueness value, $\sum_i\alpha_i$ $=$ $2\mu$ and $\sum_i \beta_i$ $=$
$2\mu$ $+$ $1$, whence the integral over the internal coordinate space vertex
can be completed and is given by the product of propagators on the right side
represented by a triangle. The quantity $\nu(\alpha_1,\alpha_2,\alpha_3)$ is
defined to be $\pi^\mu\prod_{i=1}^3a(\alpha_i)$.

It is easy to observe that for the $\sigma\bar{\psi}\psi$ vertex of (2.1) we
have $2\alpha$ $+$ $\beta$ $=$ $2\mu$ $+$ $1$ at leading order so that in
principle the integration rule of fig. 6 can be used. However, if we recall
that there is a non-zero regularization $\Delta$ this upsets the uniqueness
condition. To proceed with the determination of the divergent graph one
instead uses the method of subtractions of \cite{2}. Since we need only the
simple pole with respect to $\Delta$ and the finite part of each divergent
graph one subtracts from the particular integral another integral which has
the same divergence structure but which can be calculated for non-zero
$\Delta$. The difference of these two integrals is $\Delta$-finite and
therefore can be computed directly by uniqueness. To determine the two loop
graphs $\Pi_{1A}$ and $\Sigma_{1A}$ which occur for $\lambda_2$ we refer the
interested reader to the elementary definition of the subtracted integrals
given in \cite{4} since the treatment of the three loop graphs is new and
will be detailed here.

First, we consider the case $\Pi_{3B}$. The subtraction we used is given in
fig. 7 where the right vertex subgraph is divergent and the subtraction is
given by removing the internal $\psi$ line to join to the external right
vertex. This graph can be computed for non-zero $\Delta$ and for $\alpha$,
$\beta$ and $\lambda$ given by their leading order values. After integrating
two chains one is left with the two loop integral
\begin{equation}
\la \tilde{\mu}, \tilde{\mu}, \widetilde{\mu-\Delta}, \tilde{\mu}, 2-\mu \ra
\end{equation}
where the general definition is given in fig. 8. To compute (4.1) we first
make the transformations $\nearrow$ and $\searrow$ in the notation of
\cite{2} which leaves the integral $\la 1, \tilde{\mu}, \tilde{\mu},
\widetilde{1-\Delta}, \mu-1 \ra$ which has been computed already in \cite{7}.
Thus overall we have
\begin{eqnarray}
\la \tilde{\mu}, \tilde{\mu}, \widetilde{\mu-\Delta}, \tilde{\mu}, 2 - \mu \ra
&=& \frac{2\pi^{2\mu}}{(\mu-1)^2\Gamma^2(\mu)} \nonumber \\
&\times& \left[ 1 - \frac{\Delta}{2}\left( 3(\mu-1)\Theta + \frac{1}{(\mu-1)}
\right)\right]
\end{eqnarray}
To determine the remaining finite part one uses fig. 6 and as an intermediate
step a temporary regularization $\delta$ has to be introduced to perform
integrations in both graphs in different orders, as in \cite{2,4,6}. Useful
in obtaining the correct answer is the result
\begin{eqnarray}
\la \tilde{\mu}, \tilde{\mu}, \tilde{\mu}, \tilde{\mu}, 2-\mu-\Delta \ra
&=& \frac{2\pi^{2\mu}}{(\mu-1)^2\Gamma^2(\mu)} \nonumber \\
&\times& \left[ 1-\frac{3(\mu-1)\Delta}{2}\left( \Theta + \frac{1}{(\mu-1)^2}
\right)\right]
\end{eqnarray}
computed in a similar fashion to (4.2). After a little algebra the sum of the
finite piece and subtracted integral yield
\begin{equation}
\Pi_{3B} ~=~ - \, \frac{2\pi^{4\mu}}{(\mu-1)^3\Gamma^4(\mu)\Delta}
\left[ 1 - \frac{\Delta(\mu-1)}{2} \left( 3\Theta + \frac{1}{(\mu-1)^2}
\right)\right]
\end{equation}
To determine $\Pi_{2B1}$ we have illustrated one of two possible subtractions
in fig. 7. The procedure is the same as that for $\Pi_{3B}$ and also makes use
of (4.2) and (4.3). We obtain
\begin{equation}
\Pi_{2B1} ~=~ - \, \frac{2\pi^{4\mu}}{(\mu-1)^3\Gamma^4(\mu)\Delta}
\left[ 1 - \Delta (\mu-1) \left( 3\Theta + \frac{2}{(\mu-1)^2} \right) \right]
\end{equation}
Finally, we note
\begin{equation}
\Sigma_{1B} ~=~ - \, \frac{2\pi^{2\mu}}{(\mu-1)\Gamma^2(\mu)\Delta} ~~,~~
\Pi_{1A} ~=~ \frac{8\pi^{2\mu}}{(\mu-1)\Gamma^2(\mu)\Delta}
\end{equation}
where the finite part is zero in both cases and (4.6) are consistent with
our choice of $\chi_1$ in the renormalization of the previous section.

\sect{Computation of finite integrals.}

The remaining higher order graphs are $\Delta$-finite and therefore do not
need to be regularized. Moreover, we need only compute them for the leading
order values of $\alpha$, $\beta$ and $\lambda$. In other models, the
higher order graphs were computed by uniqueness and we used this technique
extensively for the calculation though we had to employ some novel methods
which deserve discussion. As the integrals we need to determine involve
massless propagators one can introduce conformal changes of variables on
the internal vertices on integration. For example, one conformal transformation
is
\begin{equation}
x_\mu ~ \longrightarrow ~ \frac{x_\mu}{x^2}
\end{equation}
from which it follows that
\begin{equation}
x^2 ~ \longrightarrow ~ \frac{1}{x^2}
\end{equation}
which was used in \cite{2} to compute two loop graphs. For an integral with
fermions the analogous situation for the propagators is
\begin{equation}
\xslash ~ \longrightarrow ~ \frac{\xslash}{x^2}
\end{equation}
from which we deduce that for a fermion propagating from $x$ to $y$ where both
are internal vertices
\begin{equation}
(\xslash-\yslash) ~ \longrightarrow ~ - \, \frac{\xslash(\xslash-\yslash)
\yslash}{x^2 y^2}
\end{equation}
This latter transformation provides the starting point for computing each
integral as it allows us to carry out several integrations over the internal
vertices immediately.

For example, we consider the three loop graph of fig. 4. First, integrating
one of the unique vertices and then making a conformal transformation on the
subsequent integral, which requires (5.1)-(5.4), one ends up with the first
graph of fig. 9. The fermion trace is taken over the endpoints of the
propagator with exponent $\mu$ joining to the vertex with two bosonic
propagators and the propagator with exponent $1$. The bosonic triangle of
this graph is unique and can be replaced by a unique vertex. After integrating
several chains one is left with the two loop graph of fig. 9. The techniques
to compute two loop graphs are elementary. One performs the fermion trace
which yields a series of chains of integrals and another integral proportional
to the basic integral $ChT(1,1)$ defined in \cite{2} as $ChT(\alpha_1,
\alpha_2)$ $=$ $\la \mu-1, \alpha_1, \alpha_2, \mu-1, \mu-1 \ra$ and
evaluated as
\begin{eqnarray}
ChT(\alpha_1,\alpha_2) &=& \frac{\pi^{2\mu}a(2\mu-2)}{\Gamma(\mu-1)}
\left[ \frac{a(\alpha_1)a(2-\alpha_1)}{(1-\alpha_2)(\alpha_1+\alpha_2-2)}
\right. \nonumber \\
&+& \left. \frac{a(\alpha_2)a(2-\alpha_2)}{(1-\alpha_1)(\alpha_1+\alpha_2-2)}
\right. \nonumber \\
&+& \left. \frac{a(\alpha_1+\alpha_2-1)a(3-\alpha_1-\alpha_2)}{(\alpha_1-1)
(\alpha_2-1)}\right]
\end{eqnarray}
Thus,
\begin{equation}
\Pi_{4B} ~=~ \frac{\pi^{4\mu}}{(\mu-1)^2\Gamma^4(\mu)}
\left[ 3\Theta + \frac{1}{(\mu-1)^2} \right]
\end{equation}

For the four loop graphs of fig. 4, the conformal transformations (5.1)-(5.4)
are again the starting point. For instance, in $\Pi_{6B1}$ the location of the
exponent $(2-\mu)$ and the topological structure of the graph means that
elementary integrations quickly result in an integral of the form of fig. 10,
where there is a fermion trace over the two propagators with both exponents
$1$ and another over the remaining four propagators. In fact the unique vertex
present in fig. 10 means that one only has to compute one two loop integral.
This is again achieved by taking the fermion trace and one finds
\begin{eqnarray}
\Pi_{6B1} &=& \frac{\pi^{6\mu}a(2\mu-2)}{(\mu-1)^5\Gamma^3(\mu)}
\left[ \frac{1}{(\mu-1)} - \frac{5}{2(\mu-1)^2} - \frac{(2\mu-1)\Psi}{(\mu-1)}
\right] \nonumber \\
&+& \frac{\pi^{6\mu}a^2(2\mu-2)}{(\mu-1)^8}
\end{eqnarray}
The remaining three integrals, however, turned out to require a significant
amount of effort. We consider $\Pi_{5B2}$ first. After several integrations
one is left with the first graph of fig. 11 where again there are two fermion
traces, one of which is over the two propagators with exponents $1$. Taking
this trace explicitly yields three graphs. Two of these are equivalent after
several chain integrations and are proportional to the basic integral
$\mbox{tr}\, G(2-\mu,1)$ which was defined and evaluated in \cite{7}. The
remaining integral is equal to the second graph of fig. 11, after performing
the transformations $\leftarrow$ on the right external vertex. Again taking the
fermion trace yields two graphs which are equivalent and elementary to
compute as they are proportional to $ChT(2-\mu,1)$ and a purely bosonic
integral which is the third graph of fig. 11. It is completely finite.
However, to handle infinities which cancel in our manipulations of it, we
have introduced a temporary regulator $\delta$ in the graph, which is a
standard technique in the evaluation of such complicated integrals. It is
easy to see that each of the top and bottom vertices is one step from
uniqueness and this suggests one uses integration by parts on the internal
vertex which includes the line with exponent $(3-\mu)$. The rule for this
has been given several times in previous work such as \cite{2,27}.
Consequently, one obtains the difference, after one integration, of two two
loop graphs
\begin{equation}
\la 3-\mu-\delta,\mu-1,\mu-1+\delta,1,\mu-1-\delta\ra
- \la 3-\mu,\mu-1,\mu-1+\delta,1,\mu-1-\delta\ra
\end{equation}
Since the expression is multiplied by $a(\mu-\delta)$ one needs the $O(\delta)$
term of (5.8). This is achieved by Taylor expanding each integral of (5.7) in
powers of $\delta$ but since the location of $\delta$ is common in several
exponents in each term, expanding (5.8) gives
\begin{equation}
\delta \left. \left[ \frac{\partial ~}{\partial \epsilon} ChT(3-\mu-\epsilon,
1) \right] \right|_{\epsilon=0}
\end{equation}
which can now be easily evaluated. Collecting terms and setting $\delta$ to
zero the third graph of fig. 11 is equivalent to
\begin{eqnarray}
&& \frac{\pi^{3\mu}a(2\mu-2)}{(\mu-2)^2\Gamma^2(\mu-1)} \left[ a(3-\mu)a(\mu-1)
\left( \Phi +\Psi^2 - \frac{1}{2(\mu-1)^2} \right. \right. \nonumber \\
&& + \left. \left. \frac{2}{(2\mu-3)} - 2\Psi \left( \frac{1}{2\mu-3}
+ \frac{1}{\mu-2} - \frac{1}{2(\mu-1)} \right) + \frac{2}{(2\mu-3)(\mu-2)}
\right. \right. \nonumber \\
&& - \left. \left. \frac{1}{(2\mu-3)(\mu-1)}
- \frac{1}{(\mu-2)(\mu-1)} \right) + \frac{2a^2(1)}{(\mu-2)^2} \right]
\end{eqnarray}
where $\Phi(\mu)$ $=$ $\psi^\prime(2\mu-1)$ $-$ $\psi^\prime(2-\mu)$ $-$
$\psi^\prime(\mu)$ $+$ $\psi^\prime(1)$. This completes the steps required
to compute $\Pi_{5B2}$. The final result is
\begin{eqnarray}
\Pi_{5B2} &=& - \, \frac{\pi^{6\mu}a(2\mu-2)}{(\mu-1)^5\Gamma^3(\mu)}
\left[ \frac{(2\mu-3)}{(\mu-2)}\left( \Phi + \Psi^2 - \frac{1}{2(\mu-1)^2}
\right) \right. \nonumber \\
&-& \left. \frac{(3\mu-4)\Psi}{(\mu-1)(\mu-2)^2} + \frac{1}{(\mu-2)^2}
\right] + \frac{2\pi^{6\mu}a^2(2\mu-2)}{(\mu-1)^6(\mu-2)^2}
\end{eqnarray}

For $\Pi_{5B1}$ and $\Pi_{6B2}$ a common integral lurks within each and
deserves separate treatment. It is illustrated in fig. 12 and after the
transformation $\rightarrow$ one obtains the second integral of fig. 12
where we have again introduced a temporary regulator $\delta$ in advance of
using integration by parts on the left top internal vertex. This yields a set
of four integrals, two of which are finite and proportional to the two loop
graphs $ChT(1,3-\mu)$ and $ChT(1,1)$ and two which are divergent but they
arise in such a way that the $1/\delta$ infinity cancels ie
\begin{eqnarray}
&& \pi^\mu a(\mu-\delta)a(2\mu-3)a^2(1) \nonumber \\
&& \times \left[ a(1+\delta)ChT(1,\mu-1-\delta)
- \frac{a(1)a(\mu-1+\delta)}{a(\mu-1)} ChT(1-\delta,\mu-1)\right] \nonumber \\
\end{eqnarray}
The finite part of (5.12) can easily be deduced by Taylor expanding each two
loop integral so that overall the sum of contributions to the integral of fig.
12 means that it is
\begin{eqnarray}
&& \frac{(2\mu-3)\pi^{3\mu}a^3(1)a^2(2\mu-2)}{2(\mu-2)}
\left[ 6\Theta + \frac{13}{2(\mu-1)^2} - \Phi - \Psi^2 \right. \nonumber \\
&& \left. - \, \frac{2}{(2\mu-3)^2} + \frac{1}{(2\mu-3)(\mu-1)}
+ \frac{\Psi}{(2\mu-3)(\mu-1)} \right]
\end{eqnarray}
This is the hardest part of $\Pi_{5B1}$ and $\Pi_{6B2}$ to compute. The
remaining pieces of each can easily be reduced to two loop integrals which can
be determined by methods we have already discussed. The final result for each
is
\begin{eqnarray}
\Pi_{5B1} &=& \frac{(2\mu-3)\pi^{6\mu}a(2\mu-2)}{2(\mu-1)^5(\mu-2)
\Gamma^3(\mu)} \left[ 6\Theta - \Phi - \Psi^2 + \frac{5}{2(\mu-1)^2}
- \frac{8}{(2\mu-3)} \right. \nonumber \\
&+& \left. \frac{1}{(2\mu-3)(\mu-1)} + \frac{\Psi}{(2\mu-3)(\mu-1)}
+ \frac{2(\mu-2)\Psi}{(\mu-1)} + \frac{(\mu-2)}{(\mu-1)^2} \right] \nonumber \\
\end{eqnarray}
and
\begin{eqnarray}
\Pi_{6B2} &=& - \, \frac{(2\mu-3)\pi^{6\mu}a(2\mu-2)}{(\mu-1)^5(\mu-2)}
\left[ \frac{\Phi}{2} + \frac{\Psi^2}{2} - \frac{3(\mu-1)}{(2\mu-3)}
\left( \Theta + \frac{1}{(\mu-1)^2} \right) \right. \nonumber \\
&+& \left. \frac{\Psi}{2(\mu-1)} - \frac{1}{4(\mu-1)^2}
+ \frac{(\mu-2)\Psi}{(2\mu-3)(\mu-1)} \right. \nonumber \\
&+& \left. \frac{(\mu-2)}{2(2\mu-3)(\mu-1)^2} + \frac{2}{(2\mu-3)(\mu-1)}
\right] - \frac{\pi^{6\mu}a^2(2\mu-2)}{(\mu-1)^7(\mu-2)}
\end{eqnarray}

\sect{Calculation of $\Pi_{1B}$.}

There remains only one integral to evaluate. As we have already recalled one
has to include the integral $\Pi_{1B}$ of fig. 1 in order to obtain the
exponent $\lambda_1$ correctly at leading order. In \cite{4,11} it was
determined at leading order in $1/N$. However, its $O(1/N)$ correction needs
to be included for $\lambda_2$ with the anomalous dimensions of $\alpha$,
$\beta$ and $\lambda$ now non-zero and we therefore define the $1/N$
expansion of the integral as
\begin{equation}
\Pi_{1B} ~=~ \Pi_{1B1} + \frac{\Pi_{1B2}}{N} + O \left( \frac{1}{N^2}
\right)
\end{equation}
and $\Pi_{1Bi}$ $=$ $O(1)$. The formalism to determine $\Pi_{1B2}$ has been
discussed extensively in \cite{11}. Basically, by using recursion relations it
is possible to rewrite the two bosonic integrals which occur in $\Pi_{1B}$
after taking the fermion trace as a sum of graphs which are finite at
$\alpha$ $=$ $\mu$ $+$ $\half \eta$, $\beta$ $=$ $1$ $-$ $\eta$ $-$ $\chi$
and $\lambda$ $=$ $\mu$ $-$ $1$ $+$ $O(1/N)$. As most of the integrals which
then occur have coefficients which are $O(1/N)$ one can write down the
contributions of these integrals when the fields have zero anomalous
dimensions. For one integral, though, $\la \alpha-1,\alpha-2,\alpha-1,
\alpha-1,\xi+2\ra$, this is not the case where $\xi$ $=$ $3$ $-$ $\mu$ $-$
$\eta$ $-$ $\chi$ $-$ $\lambda^\prime$ and we have set $\lambda$ $=$ $\mu$ $-$
$1$ $+$ $\lambda^\prime$. We now detail its evaluation. Manipulation using
recursion relations, \cite{11}, and using the transformations $\nearrow$ and
$\searrow$ in the notation of \cite{2} gives
\begin{eqnarray}
&& \la \alpha-1,\alpha-2,\alpha-1,\alpha-1,\xi+2\ra \nonumber \\
&&= \frac{a^2(\alpha-1)a(\xi+1)(2\mu-2\alpha-\xi)}{2(4\alpha+2\xi
-3\mu-1)a(2\alpha+\xi-\mu-1)} \nonumber \\
&& \times \left[ \frac{2(4\alpha+2\xi-3\mu-1)
(2\alpha+\xi-\mu-1)}{(\xi+1)(\mu-\xi-2)} \right. \nonumber \\
&& \left. \times \la \alpha-1,2\alpha+\xi-\mu-1,2\alpha+\xi-\mu,\alpha-1,
2\mu-2\alpha-\xi+1\ra \right. \nonumber \\
&& + \left. \left( \frac{(3\mu-4\alpha-\xi+2)(4\alpha+\xi-2\mu-3)}
{(\xi+1)(\mu-\xi-2)} - 1 \right) \right. \nonumber \\
&& \times \left. \la \alpha-1,2\alpha+\xi-\mu-1,2\alpha+\xi-\mu-1,\alpha-1,
2\mu-2\alpha-\xi+1\ra \right] \nonumber \\
\end{eqnarray}
The coefficient of each of the two integrals can easily be expanded in powers
of $1/N$, whilst the integrals themselves need to be expanded to the same
order. For the latter integral this is achieved by rewriting it as
\begin{eqnarray}
&&\left[\la \mu-1+\half\eta,\mu-1+\half\eta,1-\half\eta,1-\half\eta,2-\chi
-\lambda\ra \right. \nonumber \\
&-&\left. \la \mu-1+\half\eta,\mu-1+\chi+\lambda-\half\eta,1-\half\eta,
1-\half\eta, 2-\chi -\lambda\ra \right] \nonumber \\
&+&\la \mu-1+\half\eta,\mu-1+\chi+\lambda-\half\eta,1-\half\eta,1-\half\eta,
2-\chi -\lambda\ra
\end{eqnarray}
which is an exact result where we have first of all made a conformal
transformation based on the right external vertex, \cite{2}, followed by
mapping the integral to momentum space. However, in the second term there is
uniqueness at one of the internal vertices of integration which means it can
be computed using fig. 5 exactly and then expanded to $O(1/N)$. For the first
two terms of (6.3), since we have a difference it is easy to see that this will
be $O(1/N)$.  In other words we have chosen to subtract off an integral whose
leading order value coincides with that of the integral we require, in much the
same way as the method of subtractions is used for $\Delta$ divergent graphs.
However, here both graphs have the same structure as in fig. 8. In this linear
combination the exponents of all but one propagator coincide. Therefore the
first two integrals of (6.3) simply become
\begin{equation}
\frac{(\eta_1-\chi_1-\lambda_1)}{N} \left. \left[ \frac{\partial~}
{\partial \epsilon} \la \mu-1, \mu-1+\epsilon,1,1,2\ra \right]
\right|_{\epsilon=0}
\end{equation}
at leading order. The two loop integral can be deduced through a recursion
relation which gives a sum of $ChT(\alpha_1,\alpha_2)$ type integrals. We
record
\begin{eqnarray}
\la \mu-1,\mu-1+\epsilon,1,1,2\ra &=& \frac{2\pi^{2\mu}a(1)a(2\mu-2)
(2\mu-3)(\mu-3)}{(\mu-2)} \nonumber \\
&\times& \left[ 1 + \frac{\epsilon}{2}\left(\frac{1}{\mu-2} - \frac{2}{\mu-3}
-2 \right) \right. \nonumber \\
&-& \left. \frac{3(\mu-1)(\mu-2)\epsilon}{2(2\mu-3)(\mu-3)} \left( \Theta
+ \frac{1}{(\mu-1)^2} \right) \right]
\end{eqnarray}
Thus
\begin{eqnarray}
&& \la \alpha-1,2\alpha+\xi-\mu-1,2\alpha+\xi-\mu-1,\alpha-1,2\mu-2\alpha
-\xi+1\ra \nonumber \\
&&~= \frac{2\pi^{2\mu}(\mu-3)(2\mu-3)a(1)a(2\mu-2)}{(\mu-2)} \left[ 1
- \frac{\eta_1}{N} \left( \frac{(\mu-3)}{(\mu-2)} \right. \right. \nonumber \\
&&~+ \left. \left. \frac{(2\mu-1)(\mu-2)}{(\mu-1)} \left( 1 - \Psi
+ \frac{1}{(2\mu-3)} - \frac{1}{2(\mu-1)} \right) \right. \right. \nonumber \\
&&~+ \left. \left. \frac{(\mu-3)(2\mu^2-4\mu+1)}{(\mu-1)(\mu-2)}
+ \frac{(2\mu^2-4\mu+1)}{(\mu-3)} \right. \right. \nonumber \\
&&~+ \left. \left. \frac{3\mu(\mu-2)}{2(\mu-3)}
\left(\Theta + \frac{1}{(\mu-1)^2} \right) \right) \right]
\end{eqnarray}
Following a similar set of steps yields
\begin{eqnarray}
&& \la \alpha-1,2\alpha+\xi-\mu-1,2\alpha+\xi-\mu,\alpha-1,2\mu-2\alpha
-\xi+1\ra \nonumber \\
&&~= \frac{\pi^{2\mu}(\mu-2)(2\mu-3)a(1)a(2\mu-2)}{2(\mu-3)} \left[ 1
+ \frac{\eta_1}{N} \left( \frac{(2\mu-1)(\mu-2)}{(\mu-1)} \right. \right.
\nonumber \\
&&~\times \left. \left. \left( \Psi - \frac{1}{(2\mu-3)} - \frac{1}{2(\mu-2)}
+ \frac{1}{2(\mu-1)} - \frac{3}{2} \right) \right. \right. \nonumber \\
&&~- \left. \left. \frac{(2\mu^2-4\mu+1)(\mu-5)}{2(\mu-1)(\mu-3)}
- \frac{(\mu-1)(\mu-4)(2\mu-5)}{2(\mu-2)^2(\mu-3)} \right) \right]
\end{eqnarray}

The remaining amount of effort in determining $\Pi_{1B2}$ lies in simply
adding up all the $O(1/N)$ terms which we believe we have done correctly due
to the frequent cancellation of denominator factors such as $(2\mu-5)$,
$(\mu-4)$, $(\mu-3)$ and $(2\mu-3)$. The cancellation of the latter is
reassuring as its potential appearance in the final answer would have
indicated unwelcome singular behaviour in three dimensions. Overall we
obtained the relatively simple result
\begin{equation}
\Pi_{1B} ~=~ \frac{2\pi^{2\mu}}{(\mu-1)^2\Gamma^2(\mu)} \left[1 - \frac{\eta_1}
{N} \left( \frac{2}{(\mu-1)} - \frac{3\mu(2\mu-3)}{2}\left( \Theta
+ \frac{1}{(\mu-1)^2} \right) \right) \right]
\end{equation}
where we record that we have used the results
\begin{eqnarray}
&&  \la \mu-3,\mu-1,\mu-1,\mu-1,5-\mu\ra \nonumber \\
&& ~= \frac{a(5-\mu)}{a(1)a^2(2)} \left[ (\mu-3)^2ChT(1,1)
+ \frac{\pi^{2\mu}(\mu^3-10\mu^2+31\mu-31)a(1)}{(\mu-2)^3a(3-\mu)} \right]
\nonumber \\
&& \la \mu-2,\mu-1,\mu-1,\mu-2,5-\mu\ra \nonumber \\
&& ~= \frac{a(5-\mu)(\mu-2)}{a^3(2)}\left[ ChT(1,1)
- \frac{\pi^{2\mu}(\mu-1)a(1)}{(\mu-2)^3a(3-\mu)} \right] \\
&& \la \mu-2,\mu-1,\mu-2,\mu-1,5-\mu\ra \nonumber \\
&& ~= \frac{a(5-\mu)}{a^3(2)} \left[ ChT(1,1)
+ \frac{\pi^{2\mu}(2\mu^2-12\mu+19)a(2)}{(\mu-2)^3a(3-\mu)}\right] \nonumber
\end{eqnarray}
which were incorrectly given in \cite{11}, but were not required for that
work. We close our discussion of the evaluation of our graphs by noting that
the correct expressions for the integral $F(\alpha,\beta)$ defined in
\cite{7} is
\begin{eqnarray}
F(\alpha,\beta) &=& - \, \frac{\pi^{2\mu}a(\mu-1)a(2\mu-2)}{(\mu-1)^2}
[a(\alpha) a(2-\alpha) + a(\beta) a(2-\beta)] \nonumber \\
&+& \frac{(\alpha+\beta+\mu-3)(2-\alpha-\beta)}{(\mu-1)^2} ChT(\alpha,\beta)
\end{eqnarray}
which is valid for all $\alpha$ and $\beta$.

\sect{Discussion.}

The previous three sections have been devoted to the evaluation of the
integrals which appear in the formal master equation (3.7) and (3.8). It is
now a straightforward matter of substituting for the various expressions in
each equation and evaluating the $O(1/N^2)$ correction to the determinant. As
an intermediate step we note,
\begin{eqnarray}
\Pi_{2B} &=& \frac{\pi^{4\mu}}{(\mu-1)^2\Gamma^4(\mu)} \left[
\frac{8(2\mu-3)}{(\mu-2)}(\Phi+\Psi^2) - \frac{12(2\mu-1)\Theta}{(\mu-2)}
\right. \nonumber \\
&+& \left. \Psi \left( \frac{16}{(\mu-1)} + \frac{4(2\mu-3)(\mu-3)}
{(\mu-1)(\mu-2)^2}\right) + \frac{16}{(\mu-1)^2} - \frac{2}{(\mu-2)} \right.
\nonumber \\
&+& \left. \frac{10}{(\mu-1)} + \frac{2}{(\mu-2)^2}
- \frac{16}{\mu(\mu-2)^2\eta_1} \right]
\end{eqnarray}
and also record that
\begin{equation}
z_2 ~=~ \frac{\mu\Gamma^2(\mu)\eta^2_1}{2\pi^{2\mu}(\mu-1)} \left[
\frac{\mu}{(\mu-1)} + 2 + (2\mu-1)\Psi(\mu)\right]
\end{equation}
With these expressions together with the expansions of $p(\beta)$, $q(\beta)$,
$r(\alpha-1)$ and $s(\alpha-1)$ to the next to leading order values, we find
from the vanishing of the determinant of the matrix defined by (3.7) and (3.8)
at $O(1/N^2)$
\begin{eqnarray}
\lambda_2 &=& \frac{2\mu\eta^2_1}{(\mu-1)} \left[ \frac{2}{(\mu-2)^2\eta_1}
- \frac{(2\mu-3)\mu}{(\mu-2)} (\Phi + \Psi^2) \right. \nonumber \\
&+& \left. \Psi \left( \frac{1}{(\mu-2)^2} + \frac{1}{2(\mu-2)} - 2\mu^2
- \frac{3}{2} - \frac{1}{2\mu} - \frac{3}{(\mu-1)} \right) \right. \nonumber \\
&+& \left. \frac{3\mu\Theta}{4} \left( 9 - 2\mu + \frac{6}{\mu-2} \right)
+ 2\mu^2 - 5\mu - 3 + \frac{5}{4\mu} - \frac{1}{4\mu^2} \right. \nonumber \\
&-& \left. \frac{7}{2(\mu-1)} - \frac{1}{(\mu-1)^2} + \frac{1}{4(\mu-2)}
- \frac{1}{2(\mu-2)^2} \right]
\end{eqnarray}
which is an arbitrary dimensional expression for the $O(1/N^2)$ corrections
to the $\beta$-function of (2.2). It is worth recording that an
independent check on the correctness of (7.3) is that it ought to agree with
the expansion of the critical $\beta$-function slope computed explicitly
from the three loop result of (2.2). We have checked that this is indeed the
case. Further, we can deduce the value of the exponent in three dimensions as
\begin{equation}
\lambda ~=~ \frac{1}{2} - \frac{16}{3\pi^2N}
+ \frac{32(27\pi^2+632)}{27\pi^4N^2}
\end{equation}
As two independent exponents $\nu$ $=$ $1/(2\lambda)$ and $\eta$ or $\eta$ $+$
$\chi$ are now known at $O(1/N^2)$ this implies that the remaining
thermodynamic exponents of the model can be deduced through the hyperscaling
relations which were recently checked at leading order in \cite{13}.

With (7.4) we can now gain an improved estimate of the exponent $\nu$ in three
dimensions and compare with recent lattice simulations where the same exponent
is calculated for the case $N$ $=$ $8$ in our notation, \cite{14}. Thus
\begin{equation}
\nu ~=~ 1 + \frac{32}{3\pi^2N} - \frac{64(27\pi^2+584)}{27\pi^4N^2}
\end{equation}
and employing a Pad\'{e}-Borel technique widely used in improving estimates
of exponents, \cite{20,28,29}, we have
\begin{equation}
\nu ~=~ N \int_0^\infty dt \, e^{-Nt} \, \left. \left[
1 - \frac{16t}{3\pi^2} + \frac{32(27\pi^2+608)t^2}{81\pi^4}\right]^{-1}
\right|_{N\,=\,8} ~=~ 0.98
\end{equation}
Recent simulations, \cite{14}, give $\nu$ $=$ $0.98(7)$ and so (7.6) is in
excellent agreement with that Monte Carlo result. Also the exponent $2$ $-$
$\eta$ $-$ $\chi$, in our notation, has been calculated numerically as
$1.26(3)$ in \cite{14} and we record that from (2.6)-(2.9), at $N$ $=$ $8$,
we find $2$ $-$ $\eta$ $-$ $\chi$ $=$ $1.25$ again in good agreement.

We conclude by making several remarks. First, the Gross Neveu model has now
been solved at $O(1/N^2)$ The techniques we have had to employ are different
from those used to perform the analogous calculation in the $O(N)$ $\sigma$
model, due to the appearance of several four loop graphs. More importantly,
though, we have laid a substantial amount of the groundwork for performing
the same calculation for QED. Whilst this is a more complicated theory the
basic techniques to treat the integrals have been developed here.

\vspace{1cm}
\noindent
{\bf Acknowledgement.} The author thanks Leo K\"{a}rkk\"{a}inen and Pierre
Lacock for communications on their numerical results of the three
dimensional model.

\vspace{1cm}
\noindent
{\bf Note added.} Whilst in the final stages of this work we received a
preprint, \cite{30}, where $\lambda_2$ is stated and we record that it and
(7.3) are in agreement. We believe the method of \cite{30} is different from
the one given here.

\newpage
\noindent
{\Large {\bf Figure Captions.}}
\begin{description}
\item[Fig. 1.] Skeleton Dyson equations to determine $\lambda_1$.
\item[Fig. 2.] Higher order graph for $\psi$ consistency equation.
\item[Fig. 3.] Divergent higher order corrections to $\sigma$ consistency
equation.
\item[Fig. 4.] Finite higher order graphs for $\sigma$ consistency equation.
\item[Fig. 5.] Uniqueness rule for a bosonic vertex.
\item[Fig. 6.] Uniqueness rule for a fermionic vertex.
\item[Fig. 7.] Subtractions for $\Pi_{2B1}$ and $\Pi_{3B}$.
\item[Fig. 8.] Definition of $\la \alpha_1, \alpha_2, \alpha_3, \alpha_4,
\alpha_5 \ra$ and $\la \tilde{\alpha}_1, \tilde{\alpha}_2, \tilde{\alpha}_3,
\tilde{\alpha}_4, \tilde{\alpha}_5 \ra$.
\item[Fig. 9.] Intermediate integrals in the evaluation of $\Pi_{4B}$.
\item[Fig. 10.] $\Pi_{6B1}$ after several integrations.
\item[Fig. 11.] Intermediate integrals in the evaluation of $\Pi_{5B2}$.
\item[Fig. 12.] Common integrals in the evaluation of $\Pi_{5B1}$ and
$\Pi_{6B2}$.
\end{description}
\end{document}